\documentclass[showpacs,prl,amsmath,amssymb,superscriptaddress,nobibnotes,twocolumn, preprintnumbers]{revtex4-1}
\usepackage{graphicx}
\usepackage{endnotes}
\usepackage{bm}
\usepackage{epsfig}
\usepackage{amsmath}

\newcommand{\ie}{{\it i.e.}} 
\newcommand{\eg}{{\it e.g.}} 
\newcommand{\etal}{{\it et al.}} 

\newcommand{\WT}{WTe$_{2}$}

\newcommand{\parallelsum}{\mathbin{\!/\mkern-5mu/\!}}

\begin{document}


\title{Nearly isotropic superconductivity in layered Weyl semimetal WTe$_2$ at 98.5~kbar}

\author{Yuk~Tai~Chan}
\affiliation{Department of Physics, The Chinese University of Hong Kong, Shatin, New Territories, Hong Kong, China}

\author{P. L. Alireza}
\affiliation{Cavendish Laboratory, University of Cambridge, J. J. Thomson Avenue, Cambridge CB3 0HE, United Kingdom}

\author{K.~Y.~Yip}
\author{Q.~Niu}
\author{K.~T.~Lai}
\affiliation{Department of Physics, The Chinese University of Hong Kong, Shatin, New Territories, Hong Kong, China}

\author{Swee~K.~Goh}
\email{skgoh@phy.cuhk.edu.hk}
\affiliation{Department of Physics, The Chinese University of Hong Kong, Shatin, New Territories, Hong Kong, China}
\affiliation{Shenzhen Research Institute, The Chinese University of Hong Kong, Shatin, New Territories, Hong Kong, China}
\date{\today}


\begin{abstract}
Layered transition metal dichalcogenide WTe$_2$ has recently attracted significant attention due to the discovery of an extremely large magnetoresistance, a predicted type-II Weyl semimetallic state, and the pressure-induced superconducting state. By a careful measurement of the superconducting upper critical fields as a function of the magnetic field angle at a pressure as high as 98.5~kbar, we provide the first detailed examination of the dimensionality of the superconducting condensate in WTe$_2$. Despite the layered crystal structure, the upper critical field exhibits a negligible field anisotropy. The angular dependence of the upper critical field can be satisfactorily described by the anisotropic mass model from 2.2~K ($T/T_c\sim0.67$) to 0.03~K ($T/T_c\sim0.01$), with a practically identical anisotropy factor $\gamma\sim1.7$. The temperature dependence of the upper critical field, determined for both $H\perp ab$ and $H\parallelsum ab$, can be understood by a conventional orbital depairing mechanism. Comparison of the upper critical fields along the two orthogonal field directions results in the same value of $\gamma\sim1.7$, leading to a temperature independent anisotropy factor from near $T_c$ to $<0.01T_c$. Our findings thus identify WTe$_2$ as a nearly isotropic superconductor, with an anisotropy factor among one of the lowest known in superconducting transition metal dichalcogenides.  
\end{abstract}


\maketitle


The discovery of an extremely large and non-saturating magnetoresistance in semimetallic \WT\ \cite{Ali2014} has generated considerable research efforts \cite{Pletikosic2014, Lv2015, Cai2015, Zhu2015, Xiang2015, Jiang2015, Wang2015, Wang2016, Wu2017, Sante2017}. The interest is further intensified with the prediction that \WT\ can be a type-II Weyl semimetal, in which Weyl fermions emerge at the border between electron and hole pockets \cite{Soluyanov2015}. The crystal structure of \WT\ consists of weakly-bonded block-layers of W-Te atoms along the $c$ direction. The layered nature of \WT\ has facilitated the fabrication of devices based on thin layers of \WT, enabling the application of gate voltage, and hence further exploration of fundamental physical properties in a controllable manner \cite{Wang2015, Na2016, Yi2017, Zhang2017, Fatemi2017}.

Another powerful tool to tune the properties of \WT\ is pressure. With the application of pressure, superconductivity has been successfully induced in the bulk \WT\ \cite{Pan2015, Kang2015}. Although the temperature-pressure phase diagrams reported by two groups \cite{Pan2015, Kang2015} are quite different, some qualitative similarities can still be observed. First, the superconducting transition temperature ($T_c$) takes a dome-shaped pressure dependence, with a maximum onset $T_c$ between 6.5~K and 7~K. Second, the magnetoresistance is significantly suppressed when the superconducting state sets in. In the work of Pan \etal\ \cite{Pan2015}, superconductivity can be induced with a pressure as low as $\sim$25 kbar, which is close to the pressure range where a subtle structural transformation from $T_d$ phase to $1T'$ phase was detected via powder X-ray diffraction and Raman spectroscopy \cite{Lu2016, Zhou2016}.
However, these results contradict the study of Kang \etal\ \cite{Kang2015}, which claims that the structure of \WT\ remains the same up to 200~kbar. Despite the disagreement on the high-pressure crystal structure, the layered nature of \WT\ remains valid: the key difference between the $T_d$ phase and $1T'$ phase concerns the distinct coordination of atoms confined within the block-layer.

Given this layered nature, it is reasonable to expect an anisotropic electronic structure. However, detailed analysis of the field angle dependence of the magnetoresistance at ambient pressure reveals a surprisingly low anisotropy \cite{Thoutam2015}. If the anisotropy of the magnetoresistance is attributed to Fermi surface anisotropy, the electronic structure of \WT\ is in fact isotropic, consistent with quantum oscillations \cite{Zhu2015, Cai2015, Wu2017} and angle-resolved photoemission spectroscopy \cite{Wu2017, Sante2017} data. With the isotropic electronic structure as the backdrop, it is natural to question the dimensionality of the superconducting condensate, which motivates the present study.

The anisotropy of the upper critical field ($H_{c2}$) has provided key insight for understanding the properties of several topical superconducting systems (\eg\ Refs.~\cite{Naughton1988, Goh2012, Shimozawa2014, Mizukami2011, He2014, Yonezawa2017, Bay2012}).
In this manuscript, we report the first angular dependence of $H_{c2}$ in \WT\ at 98.5 kbar, near the pressure where $T_c$ is a maximum (onset $T_c=6$~K). Additionally, we construct and analyze the complete temperature dependence of $H_{c2}$ for both $H\perp ab$ and $H\parallelsum ab$. Our datasets allow us to probe the dimensionality of the superconductivity in \WT\ for the first time.


Single crystals of \WT\ used in this work were purchased from 2D Semiconductors. The electrical resistance ($R$) measurement was done using a standard four-probe technique on several samples cleaved from the same bulk single crystal. The electrical contacts were made with gold wires and silver paste (Dupont 6838) on freshly cleaved surfaces. Magnetic susceptibility measurement was conducted on a lump of sample consisting of multiple grains using a microcoil system \cite{Alireza2003, Goh2008}. The high pressure measurements were performed in a miniature Moissanite anvil cell similar to the one employed in Refs. \cite{Alireza2003, Goh2008, Goh2010, Alireza2017, Klintberg2012, Goh2014, Yip2017}. The culet diameter of the Moissanite anvils is 0.8~mm, and the pressure achieved was determined by the ruby fluorescence spectroscopy at room temperature. Glycerin was used as the pressure transmitting medium. The low temperature environment down to 2~K was provided by a Physical Property Measurement System (Quantum Design), and $<30$~mK by a dilution refrigerator (BlueFors Cryogenics). Both systems are equipped with a 14~T superconducting magnet. For the dilution refrigerator, a homemade rotator was used to rotate the Moissanite anvil cell in the field center of the magnet. A small Hall probe (Toshiba THS122) was glued on the anvil cell body to serve as an auxiliary sensor for the field angle. 

\begin{figure}[!t]\centering
      \resizebox{8.5cm}{!}{
              \includegraphics{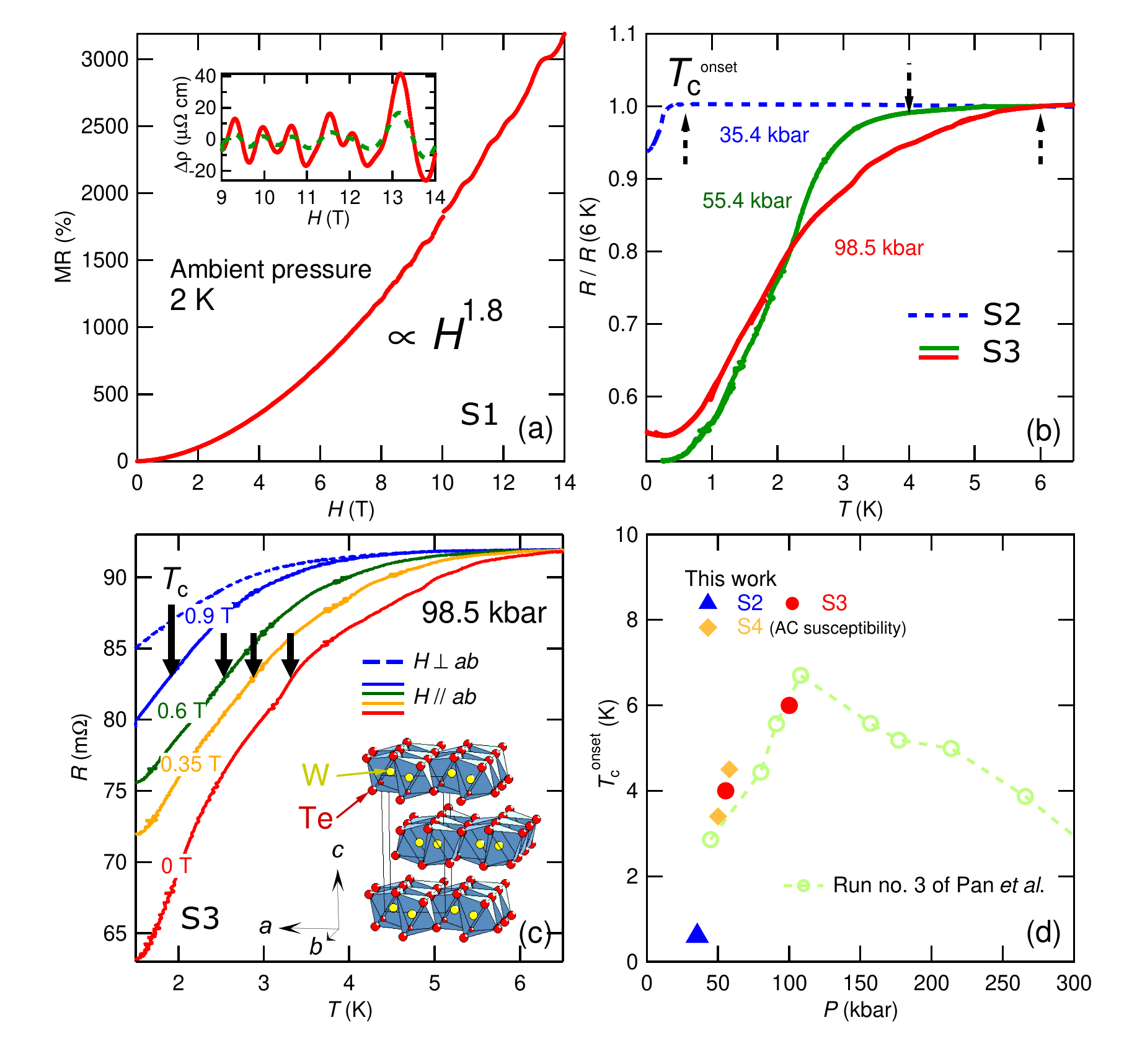}}                				
                  \caption{\label{fig1} (Color online) (a) Nearly quadratic magnetoresistance of S1 at ambient pressure, measured at $2$~K. Inset: the high-field region with background removed, showing Shubnikov-de Haas quantum oscillations at 2~K (solid line) and 6~K (dashed line). (b) Temperature-dependent resistance of samples S2 (dashed line) and S3 (solid line) at different pressures. Dashed arrows indicate the onset temperature of the superconductivity. (c) Temperature dependence of resistance for S3 at 98.5 kbar under different magnetic fields, with field applied along the $ab$-plane (dashed line) and perpendicular to the $ab$-plane (solid lines). The arrows denote the superconducting transition temperature, following the 90\% criterion. Inset: crystal structure of \WT\ at ambient pressure. (d) Pressure dependence of the superconducting onset temperature from this work (solid symbols) compared with the data of Pan \etal\ (open symbols) collected under hydrostatic conditions (Run no. 3 in \cite{Pan2015}).}
              
\end{figure}

Fig. \ref{fig1}(a) shows the magnetoresistance (MR), defined as $[R(H)-R(0)]/R(0)\times100\%$, at ambient pressure with $H\parallelsum c$ in one of the \WT\ samples (S1). The MR follows a nearly quadratic field dependence, and it reaches a magnitude of $\sim3188\%$ at 14~T and 2~K. At high field, Shubnikov-de Haas (SdH) quantum oscillations can be seen. In the inset of Fig. \ref{fig1}(a), the smooth background due to the MR is removed so that quantum oscillatory signals at 2~K, and the expected amplitude reduction at a higher temperature, can be more easily observed. At 2~K, Fourier analysis of the data gives three pronounced peaks with SdH frequencies of 97~T, 127~T and 160~T. These results are in good agreement with previous studies \cite{Ali2014, Zhu2015, Cai2015, Kang2015}, indicating good sample quality.

At high pressures, a downturn in $R$ can be observed at low temperatures (Fig. \ref{fig1}(b)). Here, two samples (S2 and S3), both cleaved from the same bulk crystal as S1, are used for high pressure studies. The onset temperature ($T_{\rm c}^{\rm onset}$) for the downturn increases as pressure is increased. At 98.5~kbar, the temperature ($T$) dependence of $R$ exhibits a clear magnetic field dependence, as shown in Fig.~\ref{fig1}(c). With an increasing field, the downturn in $R$ shifts to a lower temperature. Furthermore, $R(T)$ shows a discernible dependence on the field orientation, as indicated by the top two $R(T)$ curves, which were taken with 0.9~T applied perpendicular and parallel to the $ab$-plane, respectively. The ambient pressure structure of \WT\ is displayed as an inset of Fig.~\ref{fig1}(c), which shows the stacking of layers along the $c$ direction. The downturn is suppressed more rapidly when the field is applied perpendicular to the $ab$-plane. Additionally, we perform AC susceptibility measurement at 50~kbar and 58~kbar on S4, again cut from the same bulk crystal \cite{Note1}. The susceptibility data unambiguously prove the existence of the diamagnetic shielding due to the superconducting state. 
In Fig. \ref{fig1}(d), our $T_{\rm c}^{\rm onset}$ is plotted against pressure (solid symbols). The high-pressure data of Pan \etal\ \cite{Pan2015} collected under hydrostatic conditions, determined using the same `onset' criterion, are included for comparison. 
All these observations suggest that the downturn in $R(T)$ is associated with superconductivity, and our data are consistent with that of Pan \etal.  
The broadening of superconducting transition and the absence of the zero resistance in a similar pressure range has also been observed earlier \cite{Pan2015, Zhou2016}.
To proceed with the quantitative analysis of the superconductivity in the absence of the zero resistance, we adopt the `90\% criterion' by defining $T_c$ (upper critical field) as the temperature (field) at which the resistance is 90\% of the normal state value, as indicated by the arrows in Fig.~\ref{fig1}(c) for the case of $T_c$ (additional analyses using the 95\% criteria are provided in \cite{Note1}). 
 
\begin{figure}[!t]\centering
       \resizebox{7cm}{!}{
              \includegraphics{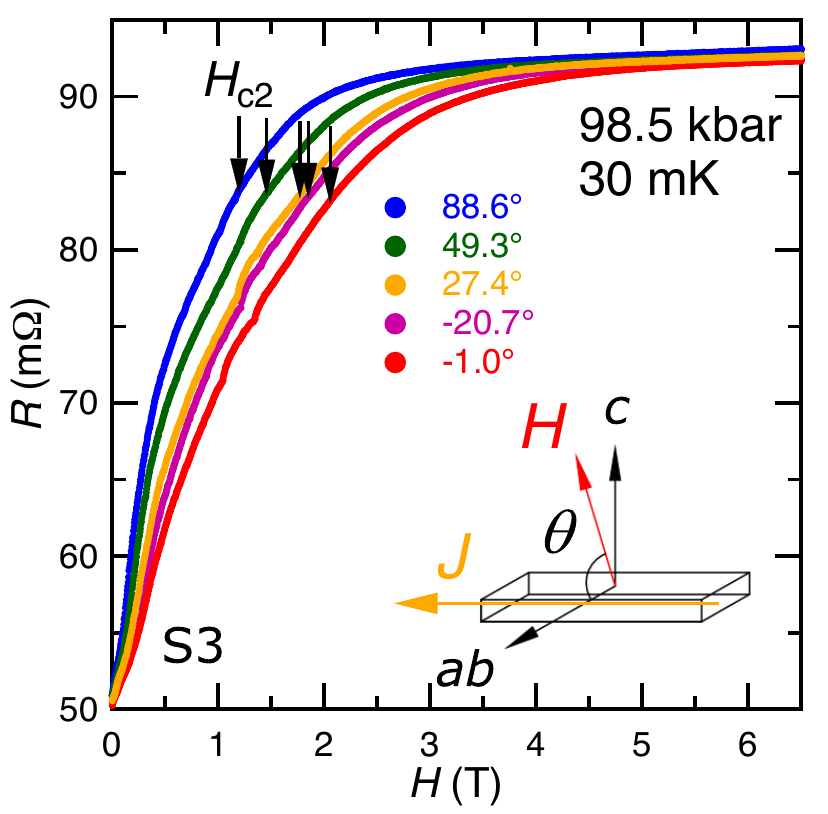}}                				
              \caption{\label{fig2} (Color online) Field dependence of resistance for S3 at $30$~mK and 98.5~kbar, at representative angles. Arrows indicate $H_{c2}$, determined using the 90\% criterion. Inset: the definition of $\theta$ with respect to the orientation of the crystal and the current direction $J$.
              }
\end{figure}

Fig. \ref{fig2} displays the field dependence of $R$ at 30~mK for S3 at 98.5~kbar. With an increasing field, $R$ increases and reaches a field independent value at a sufficiently high field. Having established the origin of the downturn in $R(T)$, the field dependence of $R$ is naturally attributed to a superconducting-to-normal states transition, with the upper critical fields ($H_{c2}$) indicated by the arrows.
We also checked that the contribution from the Hall component to $R(H)$ is negligible \cite{Note1}. Therefore, for the determination of $H_{c2}$, it is sufficient to use the positive field region of $R(H)$ only. 
With the ability to rotate the pressure cell, the magnetic field angle, $\theta$, can be varied over a large range, covering both $H\perp ab$ ($\theta=90^\circ$) and $H\parallelsum ab$ ($\theta=0^\circ$) (see the schematic drawing in Fig. \ref{fig2}).

\begin{figure}[!t]\centering
       \resizebox{8.5cm}{!}{
              \includegraphics{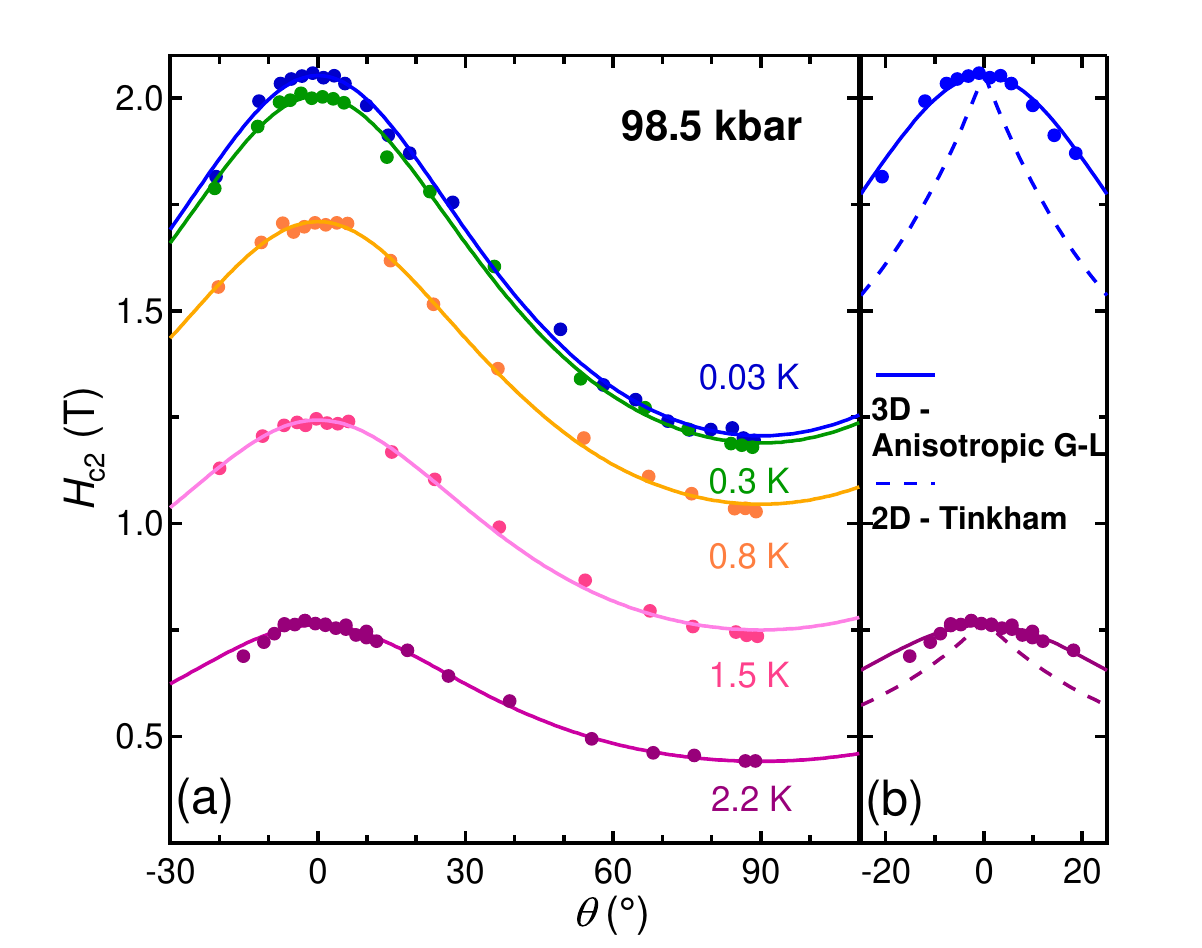}}                				
                    \caption{\label{fig3} (Color online) (a) Angular dependence of $H_{c2}$ at 0.03~K, 0.3~K, 0.8~K, 1.5~K, and 2.2~K, fitted with the 3D anisotropic Ginzburg-Landau model (solid lines). (b) Close-up of $H_{c2}(\theta)$ within $\pm25^\circ$ of the in-plane field direction. $H_{c2}(\theta)$ curves simulated using the Tinkham model with an appropriate anisotropy factor (dashed lines) are added for comparison.}
\end{figure}

Fig. \ref{fig3} presents the full angular dependence of the upper critical field, $H_{c2}(\theta)$ (symbols), over a wide temperature range, from $0.01T_c$ to $0.67T_c$. The angular dependence of $H_{c2}$ is commonly analyzed using the anisotropic mass Ginzburg-Landau (G-L) model and the Tinkham model \cite{Klemmbook, Tinkhambook, Tinkham1963, Naughton1988, Mizukami2011, Goh2012, Shimozawa2014}, which take the following form:
$$
\left[\frac{H_{c2}(\theta)\cos\theta}{H_{c2}(0^\circ)}\right]^2=1-\alpha\left|\frac{H_{c2}(\theta)\sin\theta}{H_{c2}(90^\circ)}\right|-\beta\left[\frac{H_{c2}(\theta)\sin\theta}{H_{c2}(90^\circ)}\right]^2
$$
with $(\alpha,\beta)=(1,0)$ corresponds to the Tinkham model, and $(\alpha,\beta)=(0,1)$ the anisotropic mass G-L model. Although the Tinkham model was originally developed for thin-films \cite{Tinkham1963}, \ie\ in the 2D limit, it has been successfully applied to multilayers \cite{Goh2012, Shimozawa2014}, and highly anisotropic bulk superconductors such as Bi$_{2.2}$Sr$_{1.9}$CaCu$_{2}$O$_{8+x}$ \cite{Naughton1988}. As dictated by the $|\sin\theta|$ term associated with the Tinkham model, $H_{c2}(\theta)$ would show a cusp near $\theta=0^\circ$. Our $H_{c2}(\theta)$ data vary smoothly over the entire angular range, with no evidence of a cusp-like variation near $\theta=0^\circ$. Indeed, the 3D anisotropic mass G-L model successfully describes all the $H_{c2}(\theta)$ data, as demonstrated in Fig. \ref{fig3}(a). From these fits, an important parameter can be extracted, namely the anisotropy of the upper critical fields $\gamma=H_{c2}(0^\circ)/H_{c2}(90^\circ)$: it is practically temperature independent from 0.03~K to 2.2~K, and has a small value of $1.68\pm0.05$. Fig. \ref{fig3}(b) shows the close-up of representative $H_{c2}(\theta)$ curves between $-25^\circ$ and +25$^\circ$. Using the values of $H_{c2}(0^\circ)$ and $H_{c2}(90^\circ)$, the expected $H_{c2}(\theta)$ described by Tinkham model can be simulated at each temperature. The simulations (dashed lines) clearly fail to capture the angular dependence of $H_{c2}$. All these results unambiguously point to the 3D nature of superconductivity in \WT.

\begin{figure}[!t]\centering
       \resizebox{9cm}{!}{
              \includegraphics{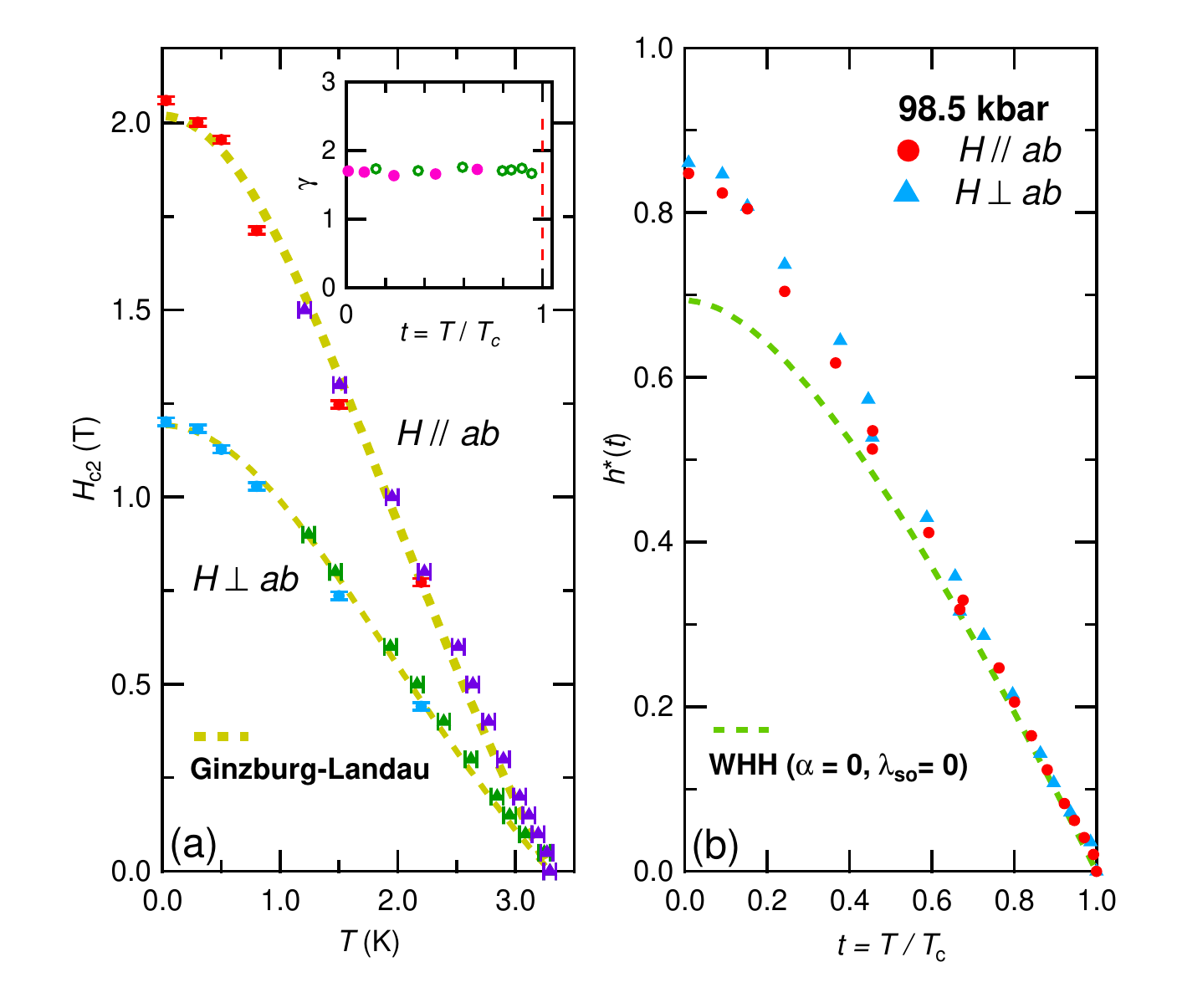}}                				
              \caption{\label{fig4} (Color online) (a) Temperature dependence of $H_{c2}$ under in-plane and out-of-plane magnetic field (symbols), fitted with Ginzburg-Landau model (dashed line). Inset: Anisotropy factor $\gamma$ against the reduced temperature $t = T/T_c$, obtained from the rotation studies (solid circles) and $H_{c2}$ data in the main panel (open circles). (b) Plot of $h^*(t)$ against $t=T/T_c$ for $H\perp ab$ (triangles) and $H\parallelsum ab$ (circles). For the definition of $h^*$, see text. The dashed line is the simulated $h^*$ using the Werthamer-Helfand-Hohenberg (WHH) theory, without spin paramagnetic or spin-orbit effects ($\alpha=0$, $\lambda_{so}=0$).}
\end{figure}

Much can be learnt by following the temperature dependence of $H_{c2}$. With the field directions carefully aligned along $H\perp ab$ and $H\parallelsum ab$, the field-temperature phase diagram of \WT\ at 98.5~kbar is constructed and plotted in Fig.~\ref{fig4}(a). The datapoints with horizontal error bars are obtained from temperature sweeps while those with vertical error bars are obtained from field sweeps, following the same 90\% criterion described earlier. The critical values, determined from both the field and temperature sweeps, exhibit an overall smooth variation. Due to the relatively low $H_{c2}$ and the availability of low temperatures, $H_{c2}(T)$ for parallel and perpendicular field directions are fully determined. The anisotropy factor $\gamma$ can be evaluated directly, shown as the open circles in the inset of Fig.~\ref{fig4}(a) for several temperatures. $\gamma$ is found to be temperature independent, and the values are in excellent agreement with the values obtained via the full fitting of $H_{c2}(\theta)$ in the framework of the 3D anisotropic mass G-L model ({\it c.f.} closed circles in the same figure). This is in stark contrast to cases where $\gamma$ exhibits a strong divergence near $T_c$, as observed in some 2D systems \cite{Mizukami2011, He2014}.

Near $T_c$, $H_{c2}$ increases linearly on cooling for both field directions, in accordance with a conventional orbital depairing behaviour. The initial slope (d$H_{c2}$/d$T$)$_{T=T_c}$ is $-0.42$~T/K and $-0.74$~T/K, for $H\perp ab$ and $H\parallelsum ab$, respectively. Note that the ratio of the slopes is 1.74, as expected. Both $H_{c2}(T)$ curves can be phenomenologically described by $H_{c2}(0)(1-t^2)/(1+t^2)$, where $t=T/T_c$ (dashed lines in Fig.~\ref{fig4}(a)). This suggests that the curves can be scaled onto each other using $\gamma$, or equivalently, the ratio of the initial slopes. Hence, the plot of $h^*(t)=H_{c2}(t)/(-$d$H_{c2}$/d$t$)$_{t=1}$ against $t$ should give a universal curve. This is indeed observed, as evidenced in Fig.~\ref{fig4}(b), indicating that the same depairing mechanism is active for both field orientations.

The Werthamer-Helfand-Hohenberg (WHH) theory \cite{Werthamer1966} is commonly applied to understand the temperature dependence of $H_{c2}$. 
The full temperature dependence of $h^*(t)$, simulated using the WHH theory for a single-band superconductor in the dirty limit, is included as the dashed line in Fig.~\ref{fig4}(b). The spin paramagnetic effect is neglected because the Pauli field estimated via $H_P(0)$[T]${=}1.8T_c$ is $\approx$6~T, far exceeds the measured $H_{c2}(0)$ in all field directions. Although the WHH theory successfully captures the variation of the upper critical field near $T_c$, it fails to describe the low temperature behaviour: the experimental $h^*(0)$ for both field directions are clearly larger than the expected WHH value of 0.69. In the clean limit, $h^*(0)=0.72$ \cite{Helfand1966}. Such a deviation from the WHH theory could be explained by, for example, a multiband effect \cite{Gurevich2007, Hunte2008}. To conclusively settle this issue, additional microscopic data at this pressure range are highly desirable.


$H_{c2}(0)$ is 1.20~T and 2.06~T for $H\perp ab$ and $H\parallelsum ab$, respectively. These values translate to an in-plane coherence length $\xi_{\parallel}=16.6$~nm, and an out-of-plane coherence length $\xi_{\perp}=9.7$~nm at the zero temperature limit. The unit cell of \WT\ contains two block layers separated by a distance $d$, stacked along the $c$ direction ({\it c.f.} inset of Fig.~\ref{fig1}c). High pressure X-ray diffraction measurement shows that the lattice constant $c$ decreases monotonically from the ambient pressure value of $\sim1.37$~nm \cite{Kang2015}. Therefore, at 98.5~kbar, we can confidently conclude that $\xi_{\perp} \gg d$. Thus, superconductivity is not confined within the block layer but rather it exhibits a non-negligible 3D character.

We now compare \WT\ with other well-known layered superconductors. In bulk Bi$_{2.2}$Sr$_{1.9}$CaCu$_{2}$O$_{8+x}$, in which $H_{c2}(\theta)$ has been demonstrated to follow the 2D Tinkham-like behaviour, $\gamma$ near $T_c$ is at least 40 \cite{Naughton1988}. On the contrary, in bulk YBa$_{2}$Cu$_{3}$O$_{7}$, which has been shown to follow the 3D anisotropic mass G-L model, $\gamma$ is $\sim8$ \cite{Naughton1988}. In other well-known transition metal dichalcogenides with $H_{c2}(\theta)$ governed by the 3D anisotropic mass G-L model, 2$H$-NbS$_{2}$ has $\gamma\approx8$ \cite{Onabe1978}, 2$H$-NbSe$_{2}$ has $\gamma>2.2$ \cite{Toyota1976, Muto1977}. Therefore, $\gamma=1.7$ is surprisingly small for a layered superconductor in which the constituent layers can be easily exfoliated from the bulk crystal, making \WT\ a superconductor with one of the lowest $\gamma$ among known superconducting transition metal dichalcogenides.

The anisotropy of the Fermi surface is directly reflected by $\gamma$. At ambient pressure, the electronic structure is isotropic with small Fermi pockets, consistent with the semimetallic nature of \WT\ \cite{Ali2014, Pletikosic2014, Lv2015, Cai2015, Zhu2015, Lu2016, Wu2017}. High-pressure Shubnikov-de Haas data up to 20~kbar detected an expansion of the Fermi pockets with increasing pressure \cite{Cai2015}. The expansion is consistent with the general trend predicted by bandstructure calculations \cite{Pan2015, Lu2016}. Crucially, bandstructure calculations find that these expanded pockets eventually touch the Brillouin zone boundary normal to the $k_z$ axis, and consequently the electronic structure acquires a substantial 2D character under pressure. According to the calculations of Lu {\it et al.} \cite{Lu2016}, at 100~kbar, the electron-like Fermi surface sheet centered at $\Gamma$ has the shape of a large oval cylinder with $k_z$ as its axis.
Although it is tricky to definitively correlate the measured $\gamma$ with the Fermi surface anisotropy of a multiband system, this Fermi surface sheet is not compatible with the observed low anisotropy factor. Thus, our work places strict constraints on the topography of Fermi surfaces, and potentially the detailed variation of the superconducting gap function, which can support superconductivity in \WT.


In summary, we have conducted a complete upper critical field study of \WT\ at 98.5~kbar, where $T_c$ is near the maximum of the dome. The angular dependence of $H_{c2}$ at all temperatures studied can be described by the 3D anisotropic mass Ginzburg-Landau model with a low anisotropy factor of $\sim1.7$. The temperature dependence of $H_{c2}$, determined for both $H\perp ab$ and $H\parallelsum ab$, can be understood by a conventional orbital depairing mechanism. The anisotropy factor, calculated directly from the ratio of $H_{c2}$ along the two orthogonal field directions, results in the same value of $\sim1.7$, leading to a temperature independent anisotropy factor from near $T_c$ to $<0.01T_c$. The measured anisotropy factor places quantitative constraints on details of Fermi surfaces and the superconducting gap function. The coherence length along the $c$ direction is much larger than the lattice parameter along the same direction. All these observations lead to the central conclusion that the pressure-induced superconductivity in layered \WT\ is nearly isotropic. On the technical side, the successful integration of the magnetic field angle with well-established extreme experimental conditions opens up a new avenue for the study of low-dimensional materials.

\begin{acknowledgments} The authors acknowledge Mike Sutherland, Jian Sun, and Shingo Yonezawa for helpful discussions. This work was supported by Research Grant Council of Hong Kong (ECS/24300214, GRF/14301316), CUHK Direct Grant (No. 3132719, No. 3132720), CUHK Startup (No. 4930048) and National Natural Science Foundation of China (No. 11504310). \end{acknowledgments}
  



 
%
%

\end{document}